\documentclass[aps,twocolumn,a4paper]{revtex4-1}
\usepackage{amsmath}
\usepackage{bm}
\usepackage{amssymb}
\usepackage{color}
\usepackage{graphicx}
\usepackage{epsfig}
\newcommand{\be}{\begin{equation}}
\newcommand{\ee}{\end{equation}}
\newcommand{\ba}{\begin{eqnarray}}
\newcommand{\ea}{\end{eqnarray}}
\newcommand{\nn}{\nonumber\\}
\begin{document}
\title{Quark and gluon distribution functions in a viscous  quark-gluon plasma medium and dilepton production via $q\bar{q}$-- annihilation} 
\author{Vinod Chandra$^{a,b}$}
\email{vchandra@iitgn.ac.in, $^{b}$: The work was started at the 
{\it Department of Physics, IISER  Bhopal, India}}
\author{V. Sreekanth$^c$}
\email{sreekanth@cts.iisc.ernet.in}
\affiliation{$^{a}$ Indian Institute of Technology Gandhinagar, 382424, Gujarat, India}
\affiliation{$^c$ Centre for High Energy Physics, Indian Institute of Science, Bangalore 560012, India}
\date{\today}
\begin{abstract}
Viscous modifications to the thermal distributions of quark-antiquarks and gluons have been 
studied in a quasi-particle description of the quark-gluon-plasma medium created in relativistic heavy-ion 
collision experiments. The model is described in terms of quasi-partons that encode the hot 
QCD medium effects in their respective effective fugacities. Both shear and bulk viscosities 
have been taken in to account in the analysis and the modifications to thermal distributions have been obtained by
modifying the the energy momentum tensor in view of the non-trivial dispersion relations
for the gluons and quarks. The interactions encoded in the equation of state induce significant modifications
to the thermal distributions. As an implication, dilepton production rate in the $q\bar{q}$ annihilation process has been investigated.
The equation of state is found to have significant impact on the dilepton production rate along with the viscosities.

\vspace{2mm}

\vspace{2mm}
\end{abstract}
\maketitle
\linespread{1.2}
\section{Introduction}
There are strong indications from relativistic heavy-ion collider
experiments (RHIC) at BNL concerning the creation of strongly coupled 
quark-gluon-plasma(QGP)~\cite{expt} that possess near perfectly 
fluidity. These observations on the QGP are mainly corroborated
by two of the most striking finding of the RHIC, {\it viz.}, the large 
elliptic flow shown by QGP, and the large jet quenching~\cite{expt} at RHIC.
The former led to the near perfect fluid picture and latter indicated
towards the strongly coupled picture of the QGP. Preliminary results from 
heavy-ion collisions at the LHC~\cite{alice,alice1} reconfirm similar picture
 of the QGP. There are interesting possibilities 
for observing the other higher order flow parameters (dipolar and triangular etc.)
at LHC, that are crucial for the quantitative understanding of collectivity and
the viscous coefficients of the QGP~\cite{higher_flow1,higher_flow2}.

The strongly coupled picture of the QGP 
is seen  to be consistent  with the lattice simulations of the QCD equation of state (EoS)~\cite{leos_lat,cheng,leos1_lat}.
The EoS is an important quantity that plays crucial role in deciding the bulk and transport properties of the QGP.
Therefore, it need to be implemented in an appropriate way as a model for the equilibrium state of the QGP
while investigating its properties within  the framework of semi-classical transport theory. Furthermore, the 
form of local thermal distribution functions that describe the hydrodynamic expansion of the 
QGP liquid must contain the effect from the realistic EoS. This sets the motivation for the present 
investigations. For the temperatures higher than the  QCD transition temperature, $T_c$, 
this issue can be addressed  by adopting the quasi-particle approaches~\cite{quasi_eff,pesh,polya,casto}.
The way to couple the non-trivial dispersion (single particle energy) for the effective degrees of freedom of 
the QGP in those quasi-particle approaches to the transport theory, is the modification 
in the definition of the $T^{\mu\nu}$~\cite{dusling,chandra_2012}.

On the other hand, transport coefficient of the QGP (shear viscosity, $\eta$ and bulk viscosity $\zeta$) are essential to 
understand and characterizes its liquid state, and the hydrodynamic evolution in heavy-ion collisions.
A tiny value of $\eta/{\mathcal S}$ (${\mathcal S}$: entropy density) can be associated with the near perfect fluid picture 
and the strongly coupled 
nature of the QGP provided that the $\zeta/{\mathcal S}$ is relatively smaller. Theoretical investigations 
suggest that this is true for the temperatures not very close to $T_c$ where bulk viscosity is 
large~\cite{kharzeev,chandra_bulk,chandra_2012}. Several phenomenological and theoretical investigations 
do suggest that the QGP indeed possess a very tiny value of the $\eta/{\mathcal S}$~\cite{shrvis,bmuller,chandra_eta}.

Moreover, in certain situations, the temperature behavior
of the  $\zeta$ and/or $\eta$ lead to cavitation and it may cause the hydrodynamic evolution of the 
QGP to stop before the freeze out is actually reached~\cite{rajagopal,Bhatt:2010cy,Bhatt:2011kr,sinha}. 
Both the bulk and shear viscosities play vital role in deciding the observed 
properties of final state hadrons in the RHIC~\cite{heinz_visco}. 
Furthermore, these transport coefficients have significant impact
on the important phenomena such as heavy quark transport ~\cite{hq}, 
photon and dilepton production in heavy-ion collisions
~\cite{Bhatt:2009zg,Bhatt:2010cy,Bhatt:2011kx,dusling1,visco_dilep,recent_lep,Bhalerao:2013aha}.
All these investigations calls for an appropriate modeling of viscous modified 
thermal distribution functions of quarks and gluons in the QGP medium. Importantly, 
such modifications naturally encode hot QCD medium effects through the QGP EoS (described 
in terms of the quasi-particle approaches at high temperature).

The present analysis is devoted to obtain 
the  viscous modified thermal distributions for quarks and gluons in the QGP medium, 
within the framework of transport theory, coupling it with a recently proposed effective fugacity 
quasi-particle model~\cite{chandra_quasi}. As an implication of these 
distribution functions, the dilepton production rate via $q\bar{q}$
annihilation process is  analyzed, and significant modifications are obtained, 
as compared to those obtained by considering a viscous quark-gluon medium with non-modified 
particle distribution functions. The ideal QCD/QGP EoS refers to the system of ultra-relativistic non-interacting gas of
quarks-antiquarks and gluons (the Stefan-Boltzmann limit of hot QCD).

The paper is organized as follows. Sec. II 
deals with a recently proposed quasi-particle description of hot QCD in terms of 
effective quasi-parton distribution functions along with how it modifies the kinetic 
theory definition of the energy momentum tensor. Furthermore, the modifications to the 
thermal distributions of the quasi-particles (quasi-gluons, and quasi-quarks) in the 
presence of dissipation that is induced by shear and bulk viscosity of the QGP, are obtained
by coupling the kinetic theory with the hydrodynamic description of the QGP. 
In Sec. III, dilepton production rate is investigated employing these viscous modified 
thermal distribution functions, and interesting observation are discussed. Sec. IV articulates the 
conclusions and future directions.

\section{Viscous modification to quark and gluon thermal distribution functions}
The determination of transport properties of any fluid is subject to the matter of 
moving away from equilibrium followed by adopting either the transport theory approach or equivalently 
the field theory approach utilizing the well known Green-Kubo formulae~\cite{green-kubo}. Once these transport coefficients
such as shear and bulk viscosities are known, it is pertinent to ask what kind of modifications 
are induced to the momentum distributions of the fluid degrees of freedom

Now, to obtain the modified distribution function of quarks and gluons which describe the viscous QGP, 
firstly we need an appropriate modeling of the equilibrium state of the QGP in terms of its degrees of freedom.
To that end, we employ a recently proposed quasi-particle description of the QGP~\cite{chandra_quasi}
as a model for its equilibrium state. This is followed by the linear perturbation induced 
in terms of shear and bulk viscous effects adopting the quadratic ansatz~\cite{dusling} 
(quadratic in terms of  momentum dependence). To obtain, viscous corrections to the 
momentum distribution of quarks-antiquarks and gluon that constitute the QGP, 
kinetic theory expression for the energy momentum tensor, $T^{\mu\nu}$ needs to be equated with its hydrodynamic 
decomposition in the presence of viscosities. Let us first briefly review the quasi-particle model followed by the 
$T^{\mu\nu}$ obtained from this model.

\subsection{The quasi-particle description of hot QCD}
Let us now discuss the quasi-particle understanding of hot QCD medium effects employed 
in the present analysis, recently proposed by Chandra and Ravishankar~\cite{chandra_quasi}.
This description has been developed in the context of the recent (2+1)-lattice QCD equation of
state (lQCD EoS)~\cite{cheng} at physical quark masses. There are more recent lattice results with the improved 
actions and more refined lattices~\cite{leos1_lat}, for which we need to re-visit the model 
with specific set of lattice data specially to define the effective gluonic degrees of freedom.
This is beyond the scope of the present analysis. Henceforth, we will stick with the one set of lattice data
utilized in the model~\cite{chandra_quasi}. 

The model initiates with an ansatz that
the lQCD EoS can be interpreted in terms of non-interacting
quasi-partons having effective fugacities, $z_{g}$, $z_q$ which encode all the interaction effects, 
where $z_g$ denotes the effective gluon fugacity, and $z_q$, denotes the 
effective quark-fugacity respectively~\cite{chandra_quasi}. In this approach, the hot QCD 
medium is divided in to two sectors, {\it viz.}, the effective gluonic sector, and 
the matter sector (light quark sector, and strange quark sector). The former refers to the contribution of 
gluonic action to the pressure which also involves contributions 
from the internal fermion lines. On the other hand, latter involve interactions among quark, anti-quarks, as well as  
their interactions with gluons.  The ansatz can be translated to the form of the equilibrium distribution functions, 
$ f_{eq}\equiv \lbrace f^{g}_{eq}, f^{q}_{eq}, f^{s}_{eq} \rbrace$ (this notation will be useful later while 
writing the transport equation in both the sector in compact notations)  as follows,

\ba
\label{eq1}
f^{g,q}_{eq} &=& \frac{z_{g,q}\exp(-\beta E_p)}{\bigg(1\mp z_{g,q}\exp(-\beta E_p)\bigg)},\nn
f^{s}_{eq} &=& \frac{z_q\exp(-\beta \sqrt{p^2+m^2})}{\bigg(1+z_q\exp(-\beta \sqrt{p^2+m^2})\bigg)},
\ea
where $E_p=\vert \vec{p}\vert \equiv p$ for gluons and light quarks, and
$\sqrt{p^2+m^2}$ for strange quarks ($m$ denotes the mass of the strange quark) and $\beta=1/T$ in the natural units.
The {\it minus} sign is for 
gluons and {\it plus} sign is for quark-antiquarks. The quarks and antiquarks possess the same distribution 
functions since we are working at the zero baryon chemical potential.The determination of $f_{eq}$ 
achieved by fixing the temperature dependence of the effective fugacities $z_g$ and $z_q$ 
from the QGP EoS which in our case is the lQCD EoS (for details we refer the reader to~\cite{chandra_quasi}).
Effective fugacity in our quasi-particle model should not be confused with the 
presence of any chemical potential. It does not indicate the presence of any conserved 
quantity in the medium. Its physical significance is described below.

It is worth emphasizing that the effective fugacity is not merely a temperature
dependent parameter which encodes the hot QCD medium effects. It is very interesting and physically significant, and
can be understood in terms of effective number density of quasi-particles in hot QCD medium, and equivalently in terms 
of an effective Virial expansion~\cite{chandra_quasi}. Interestingly, its
physical significance reflects in the modified dispersion relation both in the 
gluonic and matter sector by looking at the thermodynamic relation of energy density
$\epsilon=-\partial_\beta \ln(Z)$. One thus finds that the effective fugacities modify the single quasi-parton energy
as follows,
\ba
\label{eq2}
\omega_g&=&p+T^2\partial_T ln(z_g)\nn
\omega_q&=&p+T^2\partial_T ln(z_q)\nn
\omega_s&=&\sqrt{p^2+m^2}+T^2\partial_T ln(z_q).
\ea
This leads to the new energy dispersions for gluons ($\omega_g$), light-quarks/antiquarks ($\omega_q$)
and strange quark-antiquarks, ($\omega_s$). These dispersion relations can be explicated as follows.
The second term  in the right-hand side of Eq. (\ref{eq2}), is like the gap in the energy-spectrum due to the presence of 
quasi-particle excitations. This makes the model more in the spirit of the Landau's theory of Fermi -liquids.
A detailed discussions regarding the interpretation and physical significance of $z_g$, and $z_q$ is 
discussed at a length in~\cite{chandra_quasi,chandra_eta}. Note, that the 
quasi-particle model is reliable for the temperatures that are higher than $T_c$, hence in this 
situation, the effects induced by the strange quark mass can be neglected (strange  quark mass is around $100 MeV$ in EoS and the model is 
reliable beyond $1.5 T_c$). In this case, 
we can describe the hot QCD EoS as a system with the effective gluons ($f^g_{eq}$) and effective quark-antiquarks ($f^q_{eq}$)
as the degrees of freedom. The effective fugacity model has further been employed to study the 
anisotropic hot QCD matter and quarkonia dissociation~\cite{chandra_aniso} and 
to study the heavy-quark drag/diffusion coefficients in the QGP medium~\cite{hq}, leading to 
significant impact of the realistic QGP equation of sate on both these important phenomena.

Note that there are other  quasi-particle descriptions of hot QCD medium effects, {\it viz.}, 
the effective mass models~\cite{pesh}, effective mass models with gluon condensate~\cite{casto}, 
quasi-particle models with Polyakov loop~\cite{polya}, along with our effective fugacity model.
Our model is fundamentally distinct from these models and the differences are discussed at a length 
in~\cite{chandra_quasi}. The major difference between our model and the effective mass models is in their
philosophy itself. The effective fugacities in our model are not the effective masses. 
However,  these can be interpreted as effective mass in some limiting case ($p<< T^2 \partial_T ln(z_{g/q})$)~\cite{chandra_quasi}.
Another substantive difference, between the two approaches can be seen in terms of group velocity, $v_{gr}$ which is not the same in two approaches
In the effective mass approaches the group velocity of quasi-particles depends upon the thermal mass parameter
$v_gr= p/\sqrt{(p2 + m(T)^2}$. In contrast, in our model the description does not touch the $v_{gr}$. One can 
alternatively interpret the effective fugacities in terms of effective mass, $m_{eff}\equiv g^\prime T$ ($g^\prime$: effective coupling)
as, $z_{g,q}\equiv \exp(-m_{eff}\vert_{g,q}/T)$. The effective coupling,  $g^\prime$ comes out to be less than unity
for $T\geq 1.3\ T_c$ (both in gluonic and quark sector)~\cite{chandra_quasi}.

\subsection{Modification to the thermal distributions}
Shear and bulk viscosities are essential to understand  space-time evolution
of the QGP during its hydrodynamic expansion. Physically, shear viscosity accounts for the entropy production during the 
anisotropic expansion of the system maintaining its volume constant, on the other hand 
bulk viscosity accounts for the entropy production while the volume of the system changes at constant rate
(isotropic expansion). Since these transport coefficients are related to the non-equilibrium properties of the fluid, 
this requires to go beyond the equilibrium modeling of the fluid within linear response theory.

The general linear response (Chapman-Enskog) formalism assumes a small perturbation of 
the thermal equilibrium distribution (considering the small perturbation around
the equilibrium distributions of the quark-antiquarks and gluons) as:

\begin{equation}
\label{eq3}
f(\vec{p}, \vec{r})=f_0(p)+\delta f.
\end{equation} 
where 
\begin{equation}
f_0(p) =\frac{z_{g,q} \exp(-\beta u^\mu p_\mu)}{(1\mp z_{g,q} \exp(-\beta u^\mu p_\mu))},
\end{equation}
 denote the local thermal equilibrium distribution function  in Eq. (\ref{eq4})
in the absence of viscous effects. The quantity, $\delta f$ is the linear perturbation which encodes the viscous effects as described below.
Here, $g$ stands for quasi-gluons and $q$ for quasi-quarks (we have also neglected the mass of the strange quark which is justified at high temperature), $u^\mu$ is the 4-velocity of the fluid and $\beta=1/T$. 
The isotropic distribution, $f_0(p)$ reduced to $f_{eq}$ in the local rest frame of the fluid (LRF).

Now, using  $T \partial f_ 0/\partial(u^\mu p_\mu)=-f_0(1\pm f_0)$, the linear perturbation $\delta f$
can be expressed as~\cite{bmuller}:
\be
\label{eq4}
\delta f\equiv f(\vec{p})-f_0 (p)=f_0(p)(1\pm f_0(p)) f_1(\vec{p}).
\ee
Here, \textit{plus} is for gluons and \textit{minus} for the quark-antiquarks. The perturbation $f_1\equiv \lbrace f_{1g}, f_{1q} \rbrace$ (combined notation for quarks and gluons) 
can be thought of as a change in the argument of $f_0$ as ($ \beta u^\mu p_\mu\rightarrow \beta u^\mu p_\mu
-f_1(\vec{p}, \vec{r}))$~\cite{bmuller}, and can be thought of as a local fugacity factor leading to following form of the near-equilibrium 
distributions:
\begin{eqnarray}
\label{eq5}
f_g(\vec{p})&=&\frac{z_{g} \exp(-\beta u^\mu p_\mu +f_{1g})}{1-z_{g} \exp(-\beta u^\mu p_\mu +f_{1g})}
\nonumber\\
f_q(\vec{p})&=&\frac{z_{q} \exp(-\beta u^\mu p_\mu +f_{1q})}{1+z_{q} \exp(-\beta u^\mu p_\mu +f_{1q})}.
\end{eqnarray}
Note that Eq. (\ref{eq4}) is obtained by expanding Eq. (\ref{eq5}) and keeping only the linear term in the perturbation, $f_1$.
Next, we discuss the energy-momentum tensor for the QGP fluid obtained from these distribution functions that is essential for 
determining the form of $f_1$ in terms of shear and the bulk viscosities.

\subsubsection{Energy-momentum tensor}
To obtain viscous modifications to the quark-antiquarks and gluon distribution functions, we need to couple the fluid dynamic description of the QGP to the 
kinetic theory description. At the point of freeze-out in heavy-ion collisions, the fluid dynamic description of QGP should smoothly change to the particle description (hardonization).  
This is understood in terms of the matching of energy momentum tensor, $T^{\mu\nu}$ in these two descriptions. The $T^{\mu\nu}$ in the QGP phase is the 
energy-stress tensor encoding shear and bulk viscous effects, should match with the kinetic theory expression (in terms of $f(\vec{p})$) at freeze-out.   
To achieve the continuity of of the $T^{\mu\nu}$, in our case, the kinetic theory definition of $T^{\mu\nu}$ needs to be revised in a way that 
it must capture the hot QCD medium effects in terms of non-trivial dispersion relations and the effective fugacities. We further have to 
satisfy the the Landau-Lifshitz (LL) conditions~\cite{landau} discussed below.

The energy density and the pressure
can be obtained in terms of quasi-gluons and quasi-quarks 
in our quasi-particle model~\cite{chandra_quasi} as,
\begin{eqnarray}
\label{eq6}
 \epsilon &=&\int \frac{d^3\vec{p}}{8\pi^3}(\nu_g  \omega_g f^{eq}_g +\nu_q \omega_q f^{eq}_q)\nonumber\\
 {\mathcal P} &=& -\frac{1}{\beta} \nu_g \int \frac{d^3\vec{p}}{8\pi^3} \ln(1-z_g \exp(-\beta p))\nonumber\\
                  &+&\frac{1}{\beta} \nu_q \int \frac{d^3\vec{p}}{8\pi^3} \ln(1+z_q \exp(-\beta p)).
\end{eqnarray}
We use the notation $\nu_g=2(N_c^2-1)$ for gluonic degrees of freedom ,
$\nu_{q}=2\times 2\times N_c\times 3$ ($N_c=3$ in the present case).

In kinetic theory $T^{\mu\nu}$ is obtained from the 
single particle momentum distributions as,
\begin{eqnarray}
\label{eq7}
T^{\mu \nu} &=& \sum_{g,q} \int \frac{d^3\vec{p}}{8 \pi^3} \frac{p^\mu p^\nu}{\omega} f (\vec{p}).
\end{eqnarray}

It is emphasized in~\cite{chandra_2012}, the above expression of 
$T^{\mu\nu}$ can not simply be utilized in the present case, since, it does not
capture  the non-trivial dispersions of quasi-particles. In other words,
the thermodynamic consistency condition is not satisfied with this expression 
of $T^{\mu\nu}$ yielding incorrect expressions for energy density and the 
pressure.
 
This issue has recently been addressed in~\cite{chandra_2012} by arguing for a modified form of the 
$T^{\mu\nu}$, in a similar spirit as it is done in the effective mass
quasi-particle models~\cite{dusling}:
\begin{eqnarray}
\label{eq8}
T^{\mu\nu}&=&\sum_{g,q} \bigg\lbrace \int \frac{d^3\vec{p}}{(2\pi)^3 \omega} p^\mu p^\nu f(\vec{p})\nonumber\\
          &&+ \int \frac{d^3\vec{p}}{(2\pi)^3 p \omega} (\omega-p) p^\mu p^\nu f_0(p)\nonumber\\
          &&+ \int \frac{d^3\vec{p}}{(2\pi)^3} (\omega-p) u^\mu u^\nu f_0(p) \bigg\rbrace.
\end{eqnarray}

One can clearly realize the presence of the factors, $T^2 \frac{d ln(z_g)}{d T}$ and $T^2 \frac{d ln(z_q)}{d T}$, in the
expression for $T^{\mu\nu}$ in Eq. (\ref{eq8}) that are the part of the modified dispersions. The second term in the right-hand side of Eq. (\ref{eq8}) ensures the 
correct expression for the pressure, and the third term ensures the correct expression for the 
energy density, and hence the definition of $T^{\mu\nu}$ incorporates the thermodynamic consistency condition
correctly. In view of the reliability of the quasi-particle descriptions of hot QCD for temperature beyond
the QCD transition temperature, we may ignore the strange quark mass effects. In this case the QGP can be described 
by massless quasi-gluons, and massless quasi-quarks having non-trivial dispersion relations.
Therefore, in Eq. (\ref{eq8}), $\omega\equiv (\omega_g, \omega_q)$, and summation is over the 
gluons and quarks. On the other hand, the effective mass quasi-particle models, include the terms containing temperature derivative of the effective mass
in the modified $T^{\mu\nu}$~\cite{dusling}. It is to be noted that the effective mass models are fundamentally different from the effective fugacity 
model~\cite{chandra_2012} employed here. The differences can be understood in terms of the modified dispersions in the two cases. The major difference can be realized in terms 
of non-changing particle velocities in the effective fugacity model, in contrast, to the effective mass models. Moreover, the effective fugacity model 
can be understood in term of charge renormalization in the hot QCD, on the other hand the effective mass quasi-particle models are motivated by mass renormalization in 
the hot QCD medium.

To realize the LL conditions ($u_\mu T^{\mu\nu} u_\nu=e$ and $u_\mu \delta T^{\mu\nu} =0$),  we can resolve the $T^{\mu\nu}$ as 
\begin{equation}
\label{eq8_1}
 T^{\mu\nu}= T^{\mu\nu}_0+\delta T^{\mu\nu}.
\end{equation}
From Eq. (\ref{eq8}),
\begin{eqnarray}
 \label{eq8_2}
 T^{\mu\nu}_0&=&\sum_{g,q} \bigg\lbrace \int \frac{d^3\vec{p}}{(2\pi)^3 \omega} p^\mu p^\nu f_0(p)\nonumber\\
          &&+ \int \frac{d^3\vec{p}}{(2\pi)^3 p \omega} (\omega-p) p^\mu p^\nu f_0(p)\nonumber\\
          &&+ \int \frac{d^3\vec{p}}{(2\pi)^3} (\omega-p) u^\mu u^\nu f_0(p) \bigg\rbrace,\nonumber\\
\delta T^{\mu\nu}&=&\sum_{g,q} \int \frac{d^3\vec{p}}{(2\pi)^3 \omega} p^\mu p^\nu \delta f(\vec{p}).          
\end{eqnarray}
$T^{\mu\nu}_0$ which gets the modifications from the EoS leads to right expressions for the energy density and pressure 
following the LL condition. The form of $\delta f(\vec{p})\equiv f_0(p)(1\pm f_0(p))f_1(p)$, here is based on the quadratic ansatz  
(see Eq.(\ref{eq10})) which also follow the LL condition (since $u_\mu \pi^{\mu\nu} =0$, $u^\mu \Delta_{\mu\nu}=0$, $\Delta^{\mu\nu}:=u^\mu u^\nu-g^{\mu\nu}$).

On the other hand, the fluid dynamic definition of $T^{\mu\nu}$ in the presence of 
shear and bulk viscous effects is given as, 
\begin{equation}
\label{eq9}
T^{\mu\nu}= \epsilon u^\mu u^\nu - (p +\Pi) \Delta^{\mu\nu}+\pi^{\mu\nu},
\end{equation}
 where $\Pi$, and  $\pi^{\mu\nu}$ are the shear and bulk part of the viscous stress tensor.

 The form of the perturbations $f_1$ to the thermal distributions of gluons and quarks can be obtained in terms of the 
 $\Pi$ and $\pi^{\mu\nu}$ by relating the two definitions (kinetic theory and fluid dynamic) of the $T^{\mu\nu}$. The 
 two definitions can be matched through the following quadratic ansatz for $f_1 (\vec{p})$~\cite{dusling},
 \begin{equation}
\label{eq10}
f_1(\vec{p})=\frac{1}{(\epsilon+P) T^2} \bigg(\frac{p^\mu p^\nu}{2} C_1 \pi_{\mu\nu}
                   +\frac{C_2}{5}p^\mu p^\nu \Delta _{\mu\nu} \Pi \bigg),
\end{equation}
where the coefficients $C_1$ and $C_2$ are obtained by the matching of the two definitions of $T^{\mu\nu}$ in the 
local rest frame of the fluid (LRF). This follows from the fact that 
shear and bulk viscosities are Lorentz invariant quantities and can conveniently be obtained in the LRF of the fluid.
The factor $\epsilon+P\equiv {\mathcal S} T$ is introduced for convenience, since for the QGP in RHIC, we consider 
viscosities scaled with entropy density (${\mathcal S}$). While matching hydrodynamic and kinetic theory descriptions 
one should ensure the Landau-Lifshitz matching conditions. The modified form of the $T^{\mu\nu}$ in Eq.(\ref{eq9}) ensures  LL-matching conditions
in the temperature range where our quasi-particle model is valid~\cite{chandra_2012}.

Next, utilizing the notations in Eq. (\ref{eq4}), and matching right-hand sides of Eq. (\ref{eq8}) and  
Eq. (\ref{eq10}) in the LRF, we obtain,

\ba
\label{eq11}
\Pi \delta^{ij}+\pi^{ij}&=& \frac{\nu_{g}}{{\mathcal S} T^3}
                         \int \frac{d^3 \vec{p}}{8 \pi^3 \omega_{g} } p^i p^j p^l p^m f_g(1+ f_g)\nn
                         && \times \bigg(C_1 \pi_{lm}+\frac{C_2}{5}\Pi \bigg) \nonumber\\                  
\Pi \delta^{ij}+\pi^{ij}&=& \frac{\nu_{q}}{{\mathcal S} T^3}
                         \int \frac{d^3 \vec{p}}{8 \pi^3 \omega_{q}} p^i p^j p^l p^m f_q(1- f_q)\nn
                         && \times \bigg(C_1 \pi_{lm}+\frac{C_2}{5}\Pi\delta_{lm} \bigg).               
\ea  
Here, $l$ and $m$ are contracted and summed over. Note that shear and bulk viscous part of the stress in gluonic and quark-sector are 
distinguished by their respective transport coefficients (their total value is obtained by adding up appropriately the gluonic and quark contributions).
The fluid velocity fields for quark-antiquark and gluonic degree of freedom are assumed to be same in the QGP fluid.
The integral over the momentum in the above equations can be expressed as in~\cite{dusling}:
$I_{g,q}(\delta^{ij} \delta^{lm} +\delta^{il} \delta^{jm}+\delta^{im} \delta^{jl})$ (the subscripts $g$ and $q$
are used to distinguish the gluonic and the matter sector), where

\ba
\label{eq12}
I_g&=& \frac{1}{15 {\mathcal S} T^3} \nu_g \int \frac{d^3 \vec{p}}{8 \pi^3 \omega_{g} } p^4 f_g(1+ f_g)\nn
I_q&=&  \frac{1}{15 {\mathcal S} T^3} \nu_q \int \frac{d^3 \vec{p}}{8 \pi^3 \omega_{q} } p^4 f_q(1- f_q)
\ea
Now, from Eq. (\ref{eq11}) in the gluonic sector, 
\be
C_1 =C_2= \frac{1}{I_g},
\ee
and in the matter sector, 
\be
C_1=C_2= \frac{1}{I_q}.
\ee

The viscous modified thermal distributions of gluons and quarks in the QGP in terms of $I\equiv (I_{g,q})$,
 \be
\label{eq13}
f(\vec{p})= f_{eq}+\frac{f_{eq}(1\pm f_{eq})}{{\mathcal S} T^3} \bigg(\frac{p^\mu p^\nu}{2 I} \pi_{\mu\nu}
                  +\frac{p^\mu p^\nu \Delta _{\mu\nu} \Pi}{5 I}\bigg).
\ee
As mentioned earlier, $f_{eq}\equiv ({f_g, f_q})$.

Let us discuss the validity of the above expression of the viscous modified thermal distributions. 
The validity criterion is simply $(f-f_{eq}) << f_{eq}$ (near equilibrium condition). In other words, for the validity of our
formalism, the viscous corrections ($\pi^{\mu\nu}$ and $\Pi$) must induce small corrections to the equilibrium distribution 
of the gluons and quarks. This translates to the condition,
\be
\label{eq14}
\frac{p^\mu p^\nu \pi_{\mu\nu}}{2}+\frac{p^\mu p^\nu \delta_{\mu\nu} \Pi}{5}<< {\mathcal S} T^3 (1\pm f_{eq}) I.
\ee

Next, we consider  a case, where the integral displayed in Eq. (\ref{eq12}) can be solved analytically. 
In the limit $T^2 \partial_T (z_{g,q})/p << 1$ (high temperature limit), we can obtain analytic expressions for 
$I_g$ and $I_q$ as,
\ba
\label{eq15}
I_g&=& \frac{ 4 \nu_g T^3}{{\pi^2 \mathcal S }} PolyLog[5,z_g]\nn
I_q&=& - \frac{4 \nu_q T^3}{\pi^2 {\mathcal S}} PolyLog[5,-z_q].
\ea

\begin{figure}[h]
\includegraphics[scale=0.45]{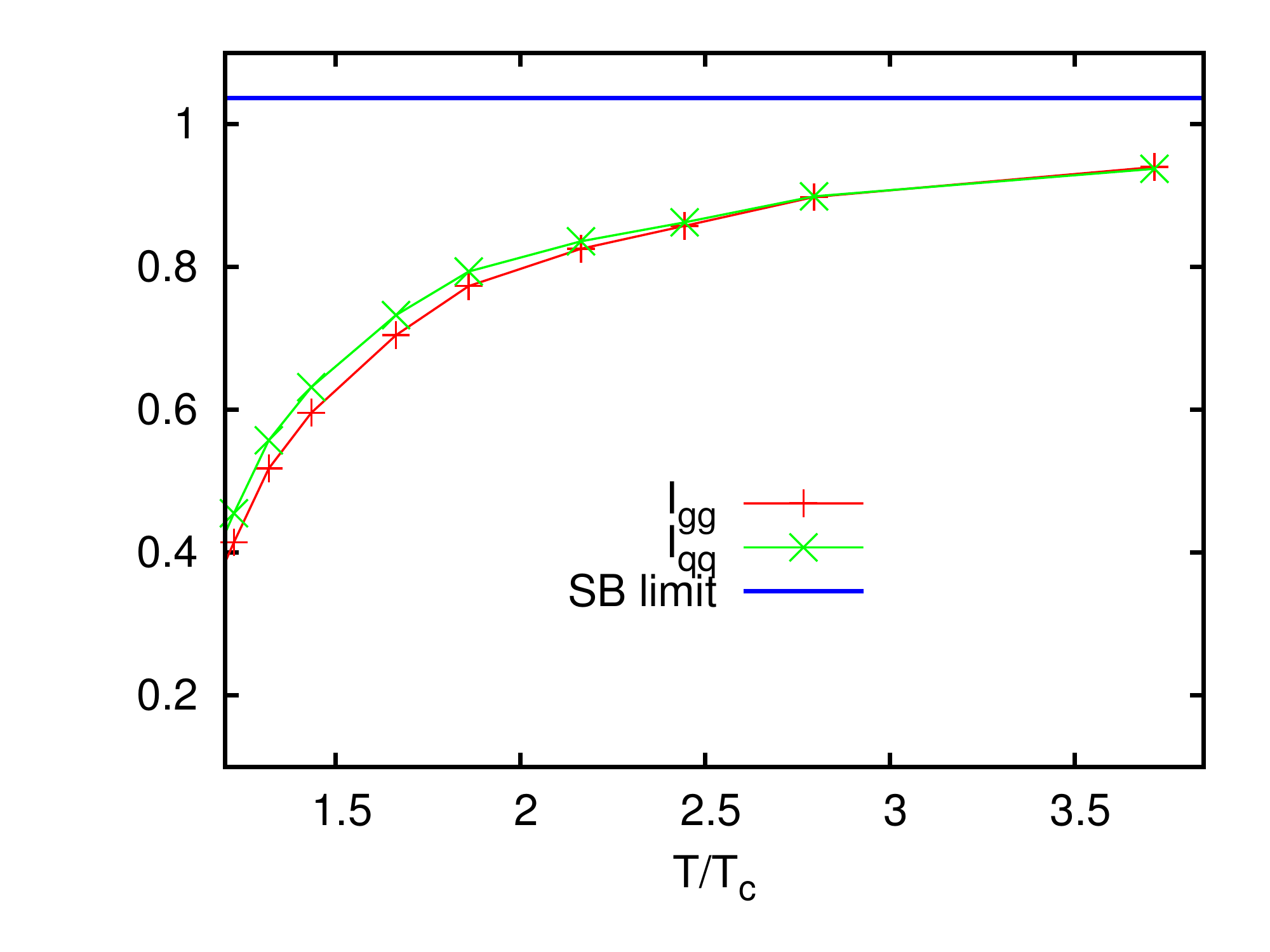}
\caption{Behavior of $I_{gg}$, and  $I_{qq}$ as function of $T/T_c$. The temperature dependence of the 
$z_g$ and $z_q$ are taken from Ref.~\cite{chandra_quasi}. Clearly the modifications induced by the EoS are 
quite significant even at higher temperatures as compared to the ideal QGP EoS.}
\label{fig1}
\end{figure}

The $PolyLog[n,x]$ function appearing in Eq. (\ref{eq15}) is having the series representation,
$PolyLog[n,x]=\sum_{k=1}^\infty \frac{x^k}{k^n}$ (convergence of the series is subject to the 
condition that $\vert x \vert \leq 1$). The Stefan-Boltzmann (SB) limit (employment of ideal QGP EoS)
is obtained only asymptotically (by putting $z_{g,q}\equiv 1$) in 
right-hand side of Eq. (\ref{eq15}). It can easily be seen that $I_g$ and $I_q$ are of the 
order of unity in the case of ideal EoS. This is also realized in~\cite{dusling}.
To see the difference in these two cases, we plot the quantities $I_{gg}$ and $I_{qq}$,
defined as:
\ba
&& I_{gg}\equiv \frac{I_g \pi^2 {\mathcal S}}{4 \nu_g T^3}=PolyLog[5,z_g]\nn
&& I_{qq}\equiv \frac{15 I_q \pi^2 {\mathcal S}}{64 \nu_g T^3}=-\frac{16}{15} PolyLog[5,-z_q],
\ea
for the ideal QGP EoS, and lQCD EoS (temperature dependence of $z_g$ and $z_q$
are taken from Ref. \cite{chandra_quasi}) in Fig. \ref{fig1}. Here, we use the identities $PolyLog[5,1]=\zeta(5)$,
and $-PolyLog[5,-1]=\frac{15}{16} \zeta(5)$ to obtain $I_g$ and $I_q$ in the case of ideal EoS.

Clearly $I_{gg}$ and $I_{qq}$, will approach to their SB limit that is $\zeta (5)$
asymptotically. The interaction effects are significant even at $3.5 T_c$. Therefore, one can not simply ignore these 
effects while obtaining the viscous modified forms of the thermal distributions of gluons and quarks in the QGP medium.
This crucial observation has been realized in the case of effective mass quasi-particle model in~\cite{dusling}.
Next, we shall investigate the significance of such viscous modified thermal distributions of gluons and quarks
in the context of dilepton production.    

\section{Effects of the EoS and viscosities on dilepton production via $q\bar{q}$ annihilation}
\label{mod-rates}
The dilepton production in the QGP medium has dominant contributions from the  $q\bar{q}$ annihilation process via the 
mechanism, $q\bar{q}\rightarrow \gamma^\ast\rightarrow l^+ l^−$. The kinetic theory expression for the 
dilepton  production  rate for a given dilepton mass and momentum is given by~\cite{Vogt},
\ba
\label{eq16}
\frac{dN}{d^4 xd^4 p}&=& \int \int \frac{d^3 \vec{p}_1}{(2\pi)^3} \frac{d^3 \vec{p}_2}{(2\pi)^3} f(E_1, T) f(E_2, T)\nonumber\\
&&\times \frac{M^2 g^2 \sigma(M^2)}{2 E_1 E_2} \delta^4 (P-p_1-p_2),
\ea
where the 4-momenta $p_{1,2} = (E_{1,2} , \vec{p}_{1,2} )$ 
are of quark and  anti-quark respectively  with $E_{1,2}=\sqrt{p_{1,2}^2 + m^2}\approx \vert\vec{p}_{1,2}\vert$, if one 
neglects the quark masses. The quantity  $M^2 = (E1 + E2 )^2 - (\vec{p}_1 + \vec{p}_2 )^2$ is the invariant mass of 
the intermediate virtual photon. Here, $g$ is the degeneracy factor, and $\sigma(M^2)$ is the thermal
dilepton production cross section. Here,  $P = p_0 = E1 + E2 ,\ \vec{p}= \vec{p}_1 + \vec{p}_2 $
is the 4- momentum of the dileptons. 

The quantity $f (E, T )$ is the quark (anti-quark) distribution function 
in thermal equilibrium, $f (E, T )=\frac{1}{1+z_q^{-1}\exp(-E/T)}$ (this form is in view of the effective quasi-particle
model based on realistic QGP EoS. In the case of ideal QGP EoS the factor $z_q$ will be replaced by unity, as done in most of
works on dilepton production in the QGP medium in the literature. As we shall argue that the EoS effects are quite significant 
even if we take high temperature limit of quark (anti-quark) distribution functions. Recall from the previous section that 
the realistic EoS strongly influence the viscous modified portion of the thermal 
distributions of gluons, and quarks (anti-quarks). In the present analysis we are interested in the  invariant masses 
that are larger compared to the temperature, $T$. In this limit, we can take the high temperature limit of 
quark (antiquark) equilibrium thermal distribution functions (replacing Fermi-Dirac
distribution with classical Maxwell-Boltzmann distribution in the case of Ideal QGP EoS) as
\be
\label{eq17}
f (E, T )\rightarrow  z_q \exp({-\frac{E}{T}}),
\ee
where $E=\vert \vec{p} \vert \equiv p$. The form will remain the same for the quarks and antiquarks since the baryon chemical potential is zero here.
It is straightforward to observe from Eq. (\ref{eq16}) that the effects coming from the EoS are of the order $z_q^2$ 
(this quantity is quite significant even at $2 T_c$). In other words, the dilepton production rate is modulated by a factor 
$z_q^2$. Let us now proceed to explore the impact of EoS and the viscous modifications to the 
dilepton production rate.

Next, we employ the result obtained in Eq. (\ref{eq13}) for the 
viscous modified quark (antiquark) distribution function $f(\vec{p})$, and take its high temperature limit, and 
analyze shear and bulk viscous contributions one by one.  In this limit, the viscous modified
quark (anti-quark) distribution functions become, 
\ba
\label{eq18}
f(\vec{p})&=& z_q\exp(-\frac{p}{T})\bigg[1 +\frac{(1-z_q\exp(-\frac{p}{T}))}{{\mathcal S} T^3} \nonumber\\ 
&& \times \bigg(\frac{p^\mu p^\nu}{2 I} \pi_{\mu\nu}
                  +\frac{p^\mu p^\nu \Delta _{\mu\nu} \Pi}{5 I}\bigg)\bigg] \nonumber\\
f(\vec{p})&\approx& z_q\exp(-\frac{p}{T}) \bigg[1 +\frac{1}{{\mathcal S} T^3}\bigg(\frac{p^\mu p^\nu}{2 I} \pi_{\mu\nu}
                +\frac{p^\mu p^\nu \Delta _{\mu\nu} \Pi}{5 I}\bigg)\bigg]\nonumber\\
\ea
Note that the first term in the above equation accounts for the equilibrium part of the quark (anti-quark)
thermal distribution, the second encodes the shear viscous effects, and the third one encodes the bulk viscous effects.

Now, the effects of viscosities on the production rate of dileptons, we employ Eq. (\ref{eq18}) to 
Eq. (\ref{eq16}), and rewrite the dilepton production rate in the component form as, 
\be
\label{eq19}
\frac{dN}{d^4 xd^4 p}= \frac{dN^{(0)}}{d^4 xd^4 p}+\frac{dN^{(\eta)}}{d^4 xd^4 p}+\frac{dN^{(\zeta)}}{d^4 xd^4 p}.
\ee

The notations $\eta$ and $\zeta$ are introduced since $\pi^{\mu\nu}$, $\Pi$ involve them as the 
first order transport coefficient in their definitions. These three terms in Eq. (\ref{eq19}) have already been 
computed for the Ideal QGP EoS in~\cite{Bhatt:2011kx}, and straight-forward 
to compute in our case (difference are there in the definition of the distribution functions). 
The first term is given by the following integral,
\begin{eqnarray}
\label{eq20}
\frac{dN^{(0)}}{d^4 xd^4 p}&=& 
\int \int \frac{d^3 \vec{p}_1}{(2\pi)^3} \frac{d^3 \vec{p}_2}{(2\pi)^3} z_q^2 exp(-\frac{E_1+E_2}{T})\nonumber\\
&&\times \frac{M^2 g^2 \sigma(M^2)}{2 E_1 E_2} \delta^4 (p-p_1-p_2)
\end{eqnarray}

This integral is  well known in the literature~\cite{Vogt} in the case of $z_q=1$.
Since $z_q$ is independent of the of the momentum of the particles, so the integral can be evaluated 
in the same way as~\cite{Vogt},
\begin{equation}
\label{eq21}
 \frac{dN^{(0)}}{d^4 xd^4 p}= \frac{z_q^2}{2} \frac{M^2 g^2 \sigma(M^2)}{2 \pi^5} exp(-\frac{p_0}{T}).
\end{equation}

The modification to rate due to the shear viscosity (at first order $\pi^{\alpha\beta}\equiv 2 \eta \sigma^{\alpha\beta}$, where 
$\sigma^{\alpha\beta}$ is the Navier-Stokes tensor ($\sigma_{\alpha\beta} =\frac{1}{2} (\nabla_\alpha u_\beta-\nabla_\beta u_\alpha)-\frac{1}{3}
\delta_{\alpha\beta}\Theta$, $\nabla_\alpha=\Delta^\mu_\alpha d_\mu$, $\Theta=d^\mu u_\mu$) can be obtained from the following equation,
\begin{eqnarray}
\label{eq22}
\frac{dN^{(\eta)}}{d^4 xd^4 p}&=& 
\int \int \frac{d^3 \vec{p}_1}{(2\pi)^3} \frac{d^3 \vec{p}_2}{(2\pi)^3} z_q^2 exp(-\frac{E_1+E_2}{T}) \nonumber\\
&& \times \frac{M^2 g^2 \sigma(M^2)}{2 E_1 E_2} \bigg[ \frac{\eta}{2 I_q {\mathcal S} T^3}(p_1^\alpha p_1^\beta+p_2^\alpha p_2^\beta)
\sigma_{\alpha\beta}\bigg] \nonumber\\
&& \times  \delta^4 (p-p_1-p_2)
\end{eqnarray}
Following the analysis of~\cite{dusling1}, we obtain the following expression for the shear viscous 
correction of the rate,

\begin{eqnarray}
\label{eq23}
\frac{dN^{(\eta)}}{d^4 xd^4 p} &=& \frac{-z_q^2}{4 \nu_q T^3 PolyLog[5,-z_q]/{\pi^2 \mathcal S}} 
\nonumber\\ &&\times \frac{1}{2} \frac{M^2 g^2 \sigma(M^2)}{(2 \pi)^5} exp(-\frac{p_0}{T}) \nonumber\\ &&\times
\frac{2}{3} \bigg[ \frac{\eta}{2 {\mathcal S} T^3} p^\alpha p^\beta \sigma_{\alpha\beta}\bigg].
\end{eqnarray}

Now, the third term which is the correction to the rate due to the bulk viscosity (at first order, $\Pi \equiv-\zeta \Theta$, where 
$\Theta$ is the expansion rate of the fluid) can be evaluated from the 
following expression,
\begin{eqnarray}
\label{eq24}
\frac{dN^{(\zeta)}}{d^4 xd^4 p}&=& 
\int \int \frac{d^3 \vec{p}_1}{(2\pi)^3} \frac{d^3 \vec{p}_2}{(2\pi)^3} z_q^2 exp(-\frac{E_1+E_2}{T}) \nonumber\\
&& \times \frac{M^2 g^2 \sigma(M^2)}{2 E_1 E_2} \bigg[ \frac{2 \zeta}{10 I_q {\mathcal S} T^3}
(p_1^\alpha p_1^\beta+p_2^\alpha p_2^\beta)
\Delta_{\alpha\beta} \Theta \bigg] \nonumber\\
&& \times  \delta^4 (p-p_1-p_2).
\end{eqnarray}

This integral can be evaluated using the analysis of~\cite{sree_33} as,
\begin{eqnarray}
\label{eq25}
\frac{dN^{(\zeta)}}{d^4 xd^4 p} &=& \frac{-z_q^2}{4 \nu_q T^3 PolyLog[5,-z_q]/{\pi^2 \mathcal S}} 
\nonumber\\ &&\times \frac{1}{2} \frac{M^2 g^2 \sigma(M^2)} {(2 \pi)^5} exp(-\frac{p_0}{T})
\\ && \times \bigg[\left(\frac{2}{3}  \frac{2 \zeta}{10 {\mathcal S} T^3} p^\alpha p^\beta \Delta_{\alpha\beta} \Theta \right)
-\frac{2}{5}\frac{\zeta}{4 {\mathcal S} T^3}M^2 \Theta \bigg].\nonumber
\end{eqnarray}

The full expression for the rate displayed in Eq. (\ref{eq20}) can be obtained by
combining Eq. (\ref{eq21}-\ref{eq25}). These expressions reduces to the those obtained in~\cite{Bhatt:2011kx}
(the expressions obtained by employing the ideal QCD EoS) by substituting $z_q=1$ (in this case $I_q\approx 1$
as already described in~\cite{dusling}).

\begin{figure}[h]
\includegraphics[scale=0.45]{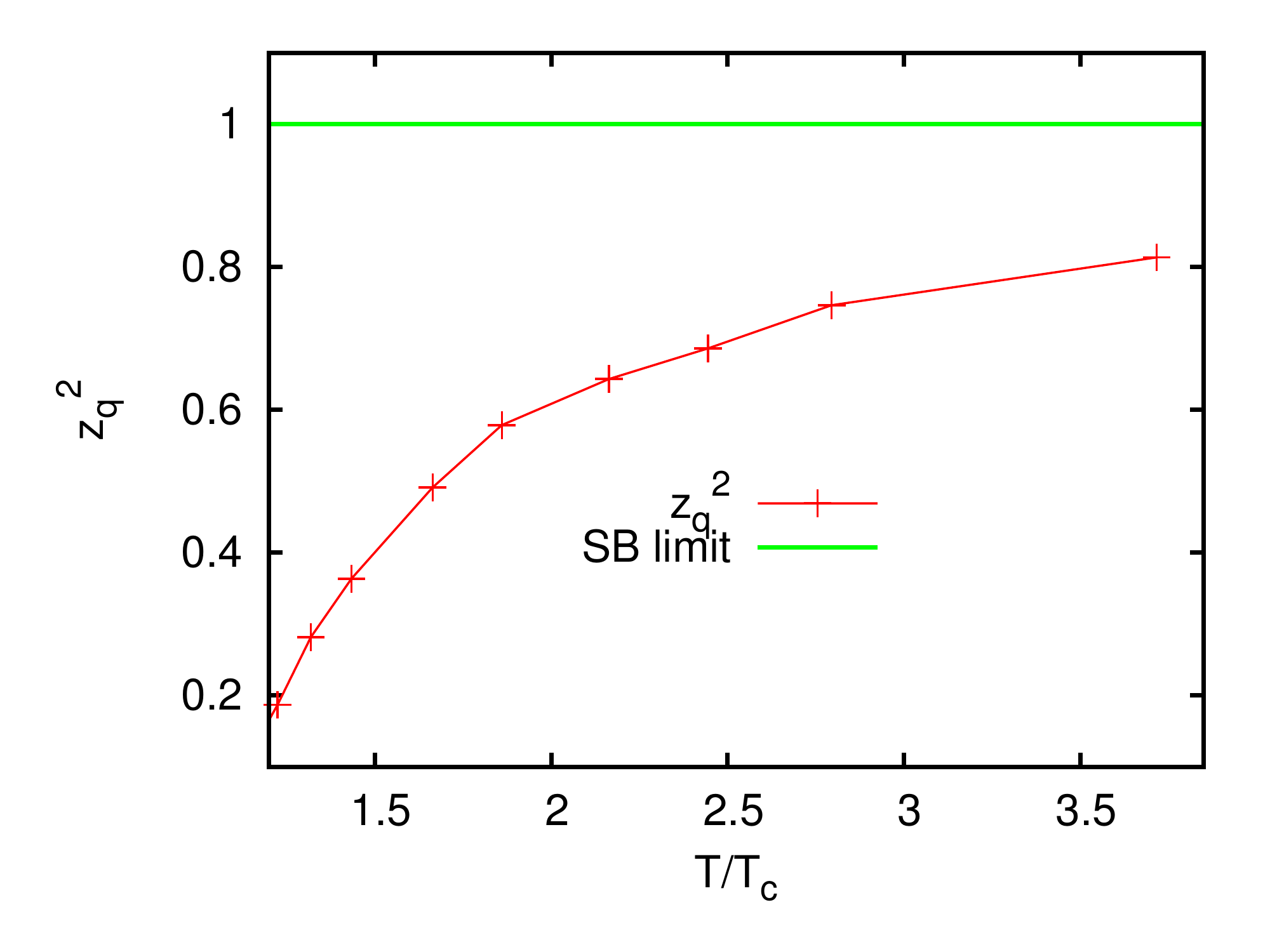}
\caption{Behavior of $z_q^2$ as function of $T/T_c$ is shown along with its SB limit ($z_q\rightarrow 1$). The temperature dependence of the 
effective quark fugacity, $z_q$ is taken from Ref.~\cite{chandra_quasi}.}
\label{fig2}
\end{figure}

\begin{figure}[h]
\includegraphics[scale=0.45]{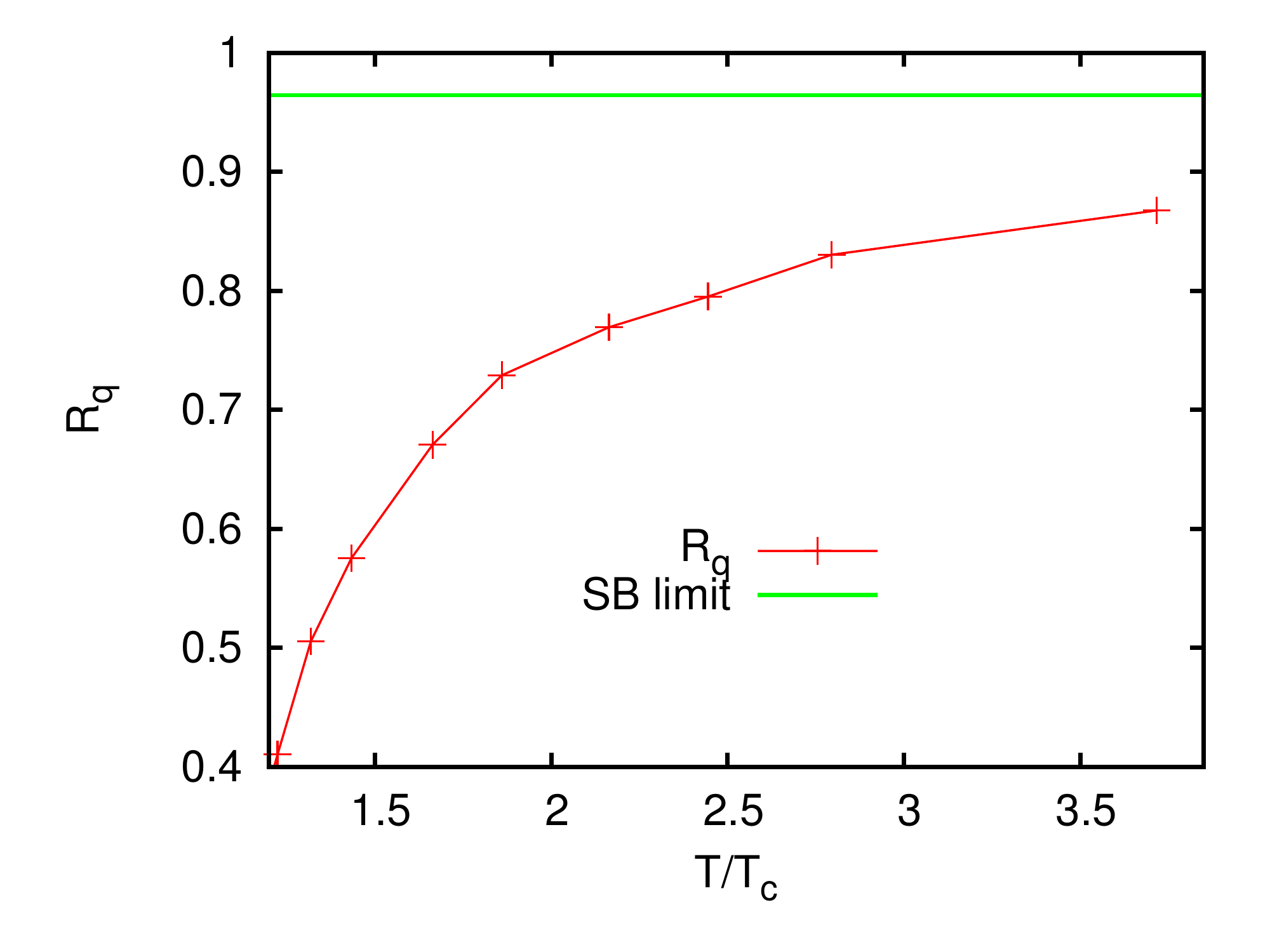}
\caption{Behavior of $R_q$ (modification factor to the viscous contribution to the dilepton rate induced by the 
realistic EoS) as function of $T/T_c$ is shown along with its SB limit ($R_q\rightarrow 1/\zeta(5)$). The temperature dependence of the 
effective quark fugacity, $z_q$ is taken from Ref.~\cite{chandra_quasi}. Here, we assume that 
$\eta$ and $\zeta$ are the phenomenological parameters for the QGP, and assumed to be same for 
realistic and ideal QGP EoSs.}
\label{fig3}

\end{figure}

If we ignore the viscous corrections, it is obvious that the EoS induced modifications appear as a factor, $z_q^2$. On the other hand, 
the shear and bulk viscous corrections to dilepton production rate gets 
a factor of $R_q=\frac{-z_q^2 \pi^2}{\nu_q PolyLog[5,-z_q]}$ (whose SB limit is $16/{15 \zeta(5)}$), as a modification 
from the EoS. We have plotted 
both of these factors, employing the quasi-particle model for (2+1)-flavor QCD~\cite{chandra_quasi} in Fig. 2 and Fig. 3.
On looking at the temperature behavior of both these factors, we can safely say that 
all the three terms in the dilepton rate in Eqs. (\ref{eq21}), (\ref{eq23}), and (\ref{eq25}) get significant modifications 
from the QGP EoS. From Figs. \ref{fig2} and \ref{fig3}, both $z_q^2$, and $R_q$ 
approach their respective SB limit only asymptotically.

Let us discuss the interesting observations that can be made out,
based on the results of the dilepton production rate obtained in 
the viscous environment and the realistic EoS. Since the quantities,  $\eta$ and $\zeta$ are the 
phenomenological numbers, hence they can safely assumed to be same in case of the ideal and the 
realistic equations of state. Therefore, the role of the viscous corrections in both the cases 
will be qualitatively  similar. However, the quantitative differences are mainly induced by the EoS.
The realization that the dilepton production is sensitive to the EoS has been also been 
seen in the recent work of Deng {\it et. al}~\cite{deng}.

Finally, the EoS induces significant modifications to the viscous modified thermal distribution functions.
These modifications play significant role in the dilepton production rate in the RHIC. The rate is 
suppressed significantly as compared to that obtained by employing the ideal EoS (the 
modifications are of the order of $z_q^2$ in the absence of the viscosities). Let us now proceed to the 
quantitative understanding of these effects from the realistic EoS and the viscosities.

We choose the value of $\eta/S=\frac{1}{4 \pi}$ (the number from the from AdS-CFT (KSS bound)~\cite{kovtun}). 
On the other hand, the bulk viscosity of QGP can taken according to the studies from
strongly interacting gauge theories~\cite{kanti} as,
\begin{equation}
\frac{\zeta}{S}=2\frac{\eta}{S} (-c_s^2+\frac{1}{3})
\end{equation}
Here, $c_s^2$ is the  speed of the sound square. The temperature dependence of $\zeta/S$ is 
dictated by $\eta/S$ and the speed of sound in the QGP phase.

\section{Dilepton spectra from heavy-ion collisions}
We now study the effect of modified gluon and quark, antiquark distribution functions 
and viscosity on thermal dilepton spectra produced from the QGP in the heavy-ion collision experiments. 
Evolution of the fireball is modeled  using relativistic hydrodynamics. In this qualitative study, we use 
one dimensional boost-invariant scaling flow to analyze the system~\cite{Bjorken:1982qr}. 
We choose the parametrization $t=\tau$ cosh$\,\eta_s$ and $z=\tau$ sinh$\,\eta_s$ for the coordinates, 
with the proper time $\tau = \sqrt{t^2-z^2}$ and space-time 
rapidity $\eta_s=\frac{1}{2}\,ln[\frac{t+z}{t-z}]$. With fluid four-velocity expressed as 
$u^\mu=(\rm{cosh}\,\eta_s,0,0,\rm{sinh}\,\eta_s)$, we can write down the equation governing the 
longitudinal expansion of the plasma as 
\begin{equation}\label{hydro}
 \frac{d\varepsilon}{d\tau}+\frac{\varepsilon + P}{\tau}=0,
\end{equation}
where we have neglected the effect of viscosity on the expansion as the significant 
contribution to particle production comes from the viscous modified rates~\cite{Bhatt:2011kx}. In order to close the 
system we use the lQCD EoS~\cite{cheng}. With critical temperature $T_C=180$ MeV, we take the 
initial conditions relevant to the RHIC energies: $\tau_0=0.5$ fm/c and $T_0=310$ MeV in our 
calculation. By numerically solving the energy equation Eq.(\ref{hydro}), we obtain the temperature profile $T(\tau)$. 
We note that for $\tau=\tau_f=6.1~$fm/c system reaches $T_c$. 

Once we obtain the temperature profile, 
particle spectra can be calculated by integrating the viscous modified 
particle production rates 
over the space-time history of the collisions
\begin{equation}
\frac{dN}{p_Tdp_TdMdy}=(4\pi M) \pi R_A^2 \int_{\tau_0}^{\tau_f}
d\tau ~\tau \int_{-y_{n}}^{y_{n}}d\eta_s
\left(\frac{1}{2}\frac{dN}{d^4xd^4p}\right).\label{yield}
\end{equation}
Where within the Bjorken model, volume element is given by 
$ d^4x=d^2x_Td\eta_s\tau d\tau=\pi R_A^2d\eta_s\tau d\tau$, with
$R_A=1.2 A^{1/3}$ representing the radius of the nucleus used for the collision (for $Au,~ A=197$). 
Here $\tau_0$ and $\tau_f$ are the initial and final values of proper time that we are interested 
(i.e. duration of the QGP phase in present analysis).
The production rates calculated in Section \ref{mod-rates} need to be modified 
while considering a longitudinally expanding system. This is done by replacing $exp(-\frac{E}{T})$ of Eq.(\ref{eq17}) 
with $exp(-\frac{u.p}{T})$ in rate expressions of Eq.(\ref{eq19}). 
With four-momentum of the dilepton parametrized as 
$p^{\alpha}$ = $(m_T coshy,p_T cos\phi_p,p_T sin\phi_p,m_Tsinhy)$, where $m_T^2$ = $p_T^2+M^2$,  
we get $u.p=m_T cosh(y-\eta_s)$. 
The other factors appearing in the rate Eqs.(\ref{eq23}) and (\ref{eq25}) are now given as
\begin{eqnarray}\label{visc-fcts1}
 p^{\alpha}p^{\beta} \sigma_{\alpha\beta} &=& \frac{2}{3\tau}p_T^2-\frac{4}{3\tau}m_T^2sinh^2(y-\eta_s),\\
 p^\alpha p^\beta \Delta_{\alpha\beta}\Theta &=& -\frac{p_T^2}{\tau}-\frac{m_T^2}{\tau}sinh^2(y-\eta_s). \label{visc-fcts2}
\end{eqnarray}
We use temperature dependent $\zeta/S$ and constant value of $\eta/S=1/4\pi$ as prescribed in the previous section
for our calculations. All results are presented for $y=0$ case only. 
\par
\begin{figure}
\includegraphics[scale=0.4]{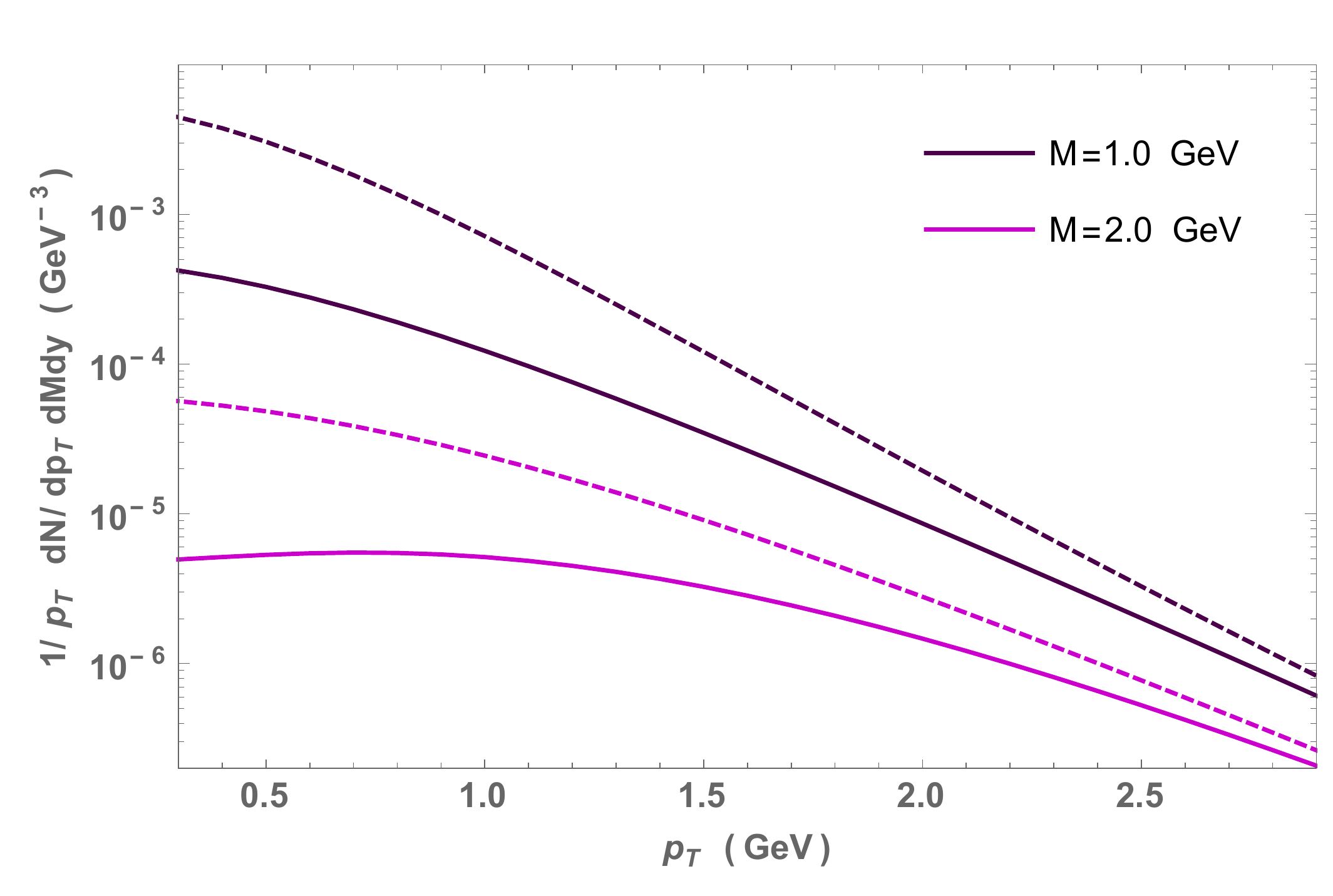}
\caption{Thermal dilepton yield from viscous quark-gluon plasma at RHIC energies for different 
dilepton invariant masses using the modified distribution functions. 
Dotted curves represent the rates calculated  ignoring the EoS modifications on the thermal distribution functions.}
\label{fig4}
\end{figure} 
We plot the transverse momentum spectra of thermal dileptons produced from the viscous QGP in Fig. \ref{fig4} for 
invariant masses $M=1,2$ GeV. In order to understand the impact of the EoS effects to the dilepton production 
through the modified distribution functions, we compare the results in a case where such effects are ignored ($z_q=1,\,I_q=1$).
Note that in both these situations, we choose the lQCD EoS for the purpose of hydrodynamical evolution of the 
temperature. Solid lines represent the rate while considering the modified distribution functions and dotted lines 
while ignoring such effects.

We observe significant modifications to the spectra while using 
the lQCD EoS through the modified distribution functions for the quark-antiquarks. 
From the curves, it is clear that effect of these terms is to suppress the particle spectra. 
Significant suppression is observed for all the values of transverse momentum e.g.; at $p_T=0.5$ GeV, 
suppression is about 89$\%$ for $M=2$ GeV, which 
reduces to about 79$\%$ for $p_T=1$ GeV and about 47$\%$ for $p_T=2$ GeV.

We also observe that suppression of low $p_T$ particles is strong indicating that 
effect of the modifications is more dominant in the later stages of collision, when system is near $T_c$. 
The high $p_T$ particles, produced predominantly during the early stages of the evolution of the 
system, are also affected by these modifications albeit less compared to the low $p_T$ region. 
This behavior can be understood by observing the behavior of the EoS induced modifications to the dilepton rate in 
Fig \ref{fig3}.
Since these terms are getting multiplied with the dilepton rates employing ideal EoS, they suppress the spectra at low $p_T$.

\begin{figure}
\includegraphics[scale=0.43]{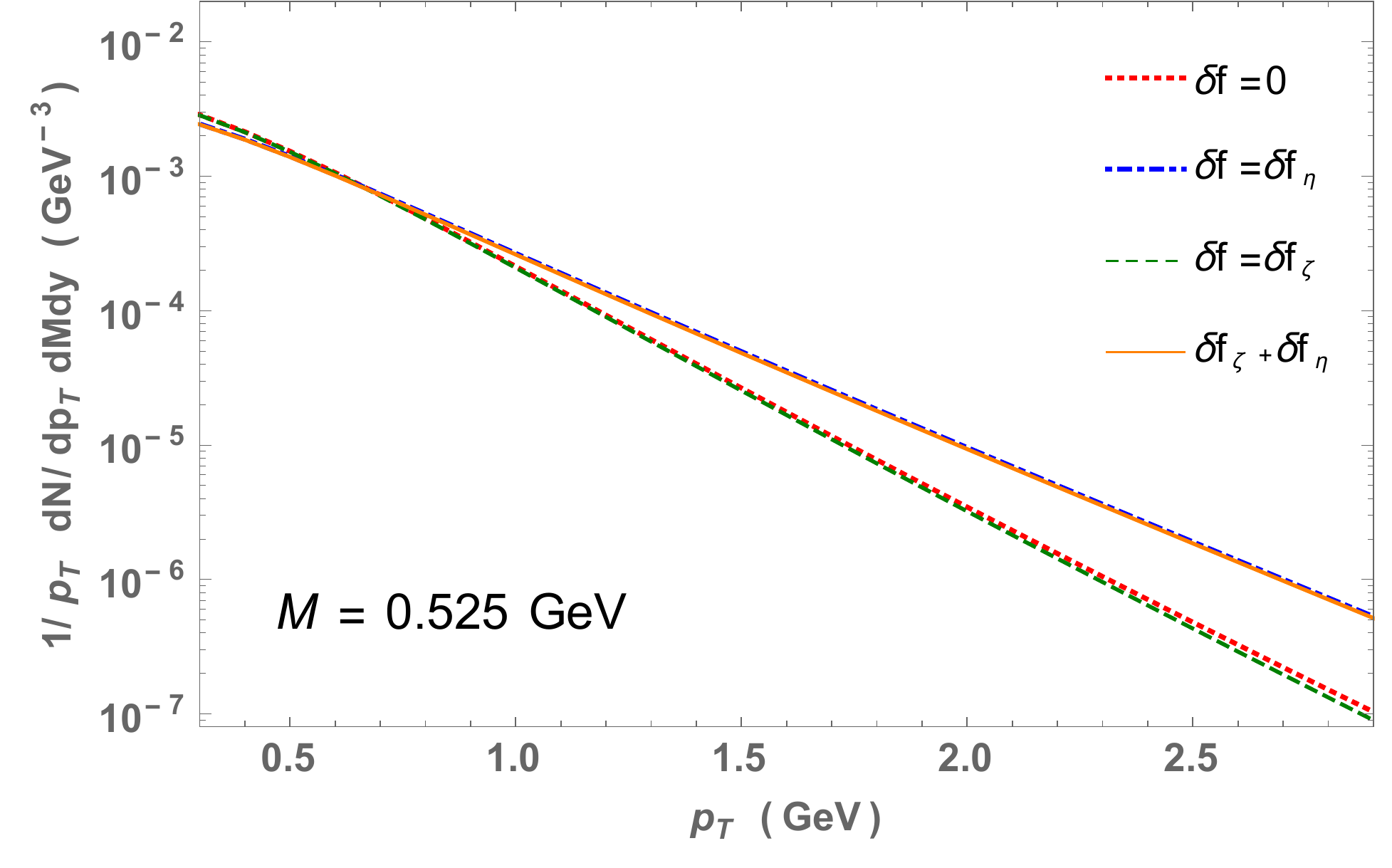}
\caption{Transverse momentum spectra for the thermal dileptons (with invariant mass $M=0.525$ GeV) from 
the viscous quark-gluon plasma produced in RHIC energies. Effect of viscosities is highlighted.}
\label{fig5}
\end{figure}

Next, we look into the behavior of various dissipative terms in the dilepton production. 
In Fig. \ref{fig5}, we plot dilepton yield as a function of transverse momentum of the pair for the 
invariant mass $M=0.525$ GeV. In this plot we show the effect of various viscosity terms in the 
total rate. Firstly,  we plot the spectra without considering the viscous effects in the calculations ($\delta f=0$ case).
As emphasized earlier, the presence of  $z_q^2$ terms in the expression, as we saw from Eq. [\ref{eq21}], will 
lead to  the overall suppression of  the spectra. One can see that inclusion of the the effect of bulk viscosity ($\delta f=\delta f_\zeta$) 
has only marginal effect. 
As observed before in~\cite{Bhatt:2011kx}, at high $p_T$, effect of bulk viscosity is to suppresses the spectra, 
for e.g.; at $p_T=2$ GeV suppression is about 7$\%$. 
Next, we consider only the effect of shear viscosity ($\delta f=\delta f_\eta$ case) in the spectra. 
Shear viscosity significantly enhances the particle production and its effect becomes stronger as $p_T$ increases. 
This can be understood 
from the presence of the first (positive) term on the right hand side of Eq.(\ref{visc-fcts1}). 
At $p_T=2$ GeV enhancement of spectra due to shear viscosity is around 177$\%$. 
Since the shear viscosity coefficient $\eta/S>\zeta/S$ in the entire temperature regime we are interested, 
its effect is expected to be more dominant. As before~\cite{Bhatt:2009zg,Bhatt:2011kx}, we observe that even the lowest value of 
shear viscosity $\sim1/4\pi$ has significant effect on the spectra. 
Consequently, when we consider fully viscous 
case ($\delta f=\delta f_\eta+\delta f_\zeta$), spectra gets highly enhanced due to 
shear viscosity, albeit the presence of bulk viscosity suppression.   
For instance, at $p_T=2$ GeV the total enhancement of the spectra is 170$\%$.

It is worth emphasizing that the main source of thermal dileptons in the QGP medium is the quark-anti-quark annihilation processes considered here. 
There are other higher order processes that can also contribute to the  thermal dilepton production~\cite{Altherr:1992th,Thoma:1997dk}.
Such higher order processes are not considered in the present analysis. It may further be noted that the thermal dileptons 
from the annihilation process is dominant in the regime of intermediate invariant mass $1< M <3$ GeV and transverse momentum 
of the pair, $p_T$ in that range~\cite{vanHees:2007ma,Akiba:2009es}. We intend to extend our present studies by incorporating higher order contributions 
in view of the quasi-particle description in the near future.

\section{Conclusions}
In conclusion, the form of  viscous modified thermal 
distribution functions for quasi-quarks and quasi-gluons 
are obtained in the QGP medium by systematically 
employing the realistic EoS for the QGP (the lQCD EoS). As an implication,
the impact of them  is demonstrated on the dilepton production via $q\bar{q}$ annihilation in RHIC.
The EoS also induces significant modifications to the viscous modified thermal distributions of the gluons and 
quark-antiquarks that constitute the QGP. The effects are equally 
significant in deciding the dilepton production rate in the viscous 
QGP medium. In particular, even in the high temperature regime, where 
the hot QCD medium effects are weaker, the realistic EoS and
viscosities play crucial role.

Finally, coupling the present analysis to the relativistic second order viscous hydrodynamic evolution of the
QGP and impact of the temperature dependence of the shear and bulk viscosities 
on the dilepton production rate will be matters of future investigation.

\section*{Acknowledgments}
VC acknowledges financial support from the Department of Science and Technology, Govt. of India [Grant:DST/IFA-13/PH-55]. 
We sincerely thank F. Becattini and V. Ravishankar for fruitful discussions and encouragement. We are highly indebted to the people of 
India for their generous support for the research in basic sciences.

\end{document}